\documentclass[reprint,aip,pof,amssymb,amsmath,superscriptaddress,showpacs,footinbib]{revtex4-1}
\usepackage{graphicx,subfigure,subeqnarray,fancyhdr,amsmath,epstopdf,multirow,color, mathrsfs}

\begin{document}

\renewcommand{\labelitemi}{$-$}
\newcommand{\change}[1]{{\color{black} #1}}
\def\v{{\vspace{2cm}}}
\newcommand{\Fc}{\mathcal{F}}\newcommand{\Rc}{\mathcal{R}}\newcommand{\dd}{\mathrm{d}}\newcommand{\ee}{\mathrm{e}}\newcommand{\ci}{\mathrm{i}}\newcommand{\ib}{\mathbf{i}}\newcommand{\jb}{\mathbf{j}}\newcommand{\kb}{\mathbf{k}}\newcommand{\ab}{\mathbf{a}}\newcommand{\Fb}{\mathbf{F}}\newcommand{\fb}{\mathbf{f}}\newcommand{\Gb}{\mathbf{G}}\newcommand{\Mb}{\mathbf{M}Ä}\newcommand{\nb}{\mathbf{n}}\newcommand{\Sb}{\mathbf{S}}\newcommand{\Sbs}{\mathbf{S^*}}\newcommand{\Rb}{\mathbf{R}}\newcommand{\Sigb}{\boldsymbol{\Sigma}}\newcommand{\Sigbs}{\boldsymbol{\Sigma^*}}\newcommand{\alphab}{\boldsymbol\alpha}\newcommand{\omegab}{\boldsymbol{\omega}}\newcommand{\epsb}{\boldsymbol{\epsilon}}\newcommand{\ub}{\mathbf{u}}\newcommand{\eb}{\mathbf{e}}\newcommand{\vv}[1]{\underline{#1}}\newcommand{\ev}{\vv{e}}\newcommand{\rv}{\vv{r}}\newcommand{\TT}[1]{\underline{\underline{#1}}}\newcommand{\omb}{\mathbf{\omega}}\newcommand{\Ub}{\mathbf{U}}\newcommand{\xb}{\mathbf{x}}\newcommand{\rb}{\mathbf{r}}\newcommand{\ssb}{\mathbf{s}}\newcommand{\Xb}{\mathbf{X}}\newcommand{\Pe}{\mbox{Pe}}\newcommand{\mean}[1]{\langle #1\rangle}\newcommand{\ddp}{[p]^\pm}\newcommand{\taub}{\mbox{\boldmath$\tau$}}\newcommand{\Fr}{\mbox{\textit{Fr}}}\let\grad\nabla\newcommand{\z}{\zeta}\newcommand{\kk}{{\cal D}}\newcommand{\tkk}{\tilde{{\cal D}}}\newcommand{\e}{\varepsilon}\newcommand{\zb}{\bar{\zeta}}\let\grad\nabla\let\bcdot\cdot\newcommand{\half}{{\textstyle\frac{1}{2}}}\newcommand{\textfrac}[2]{{\textstyle\frac{#1}{#2}}}\newcommand{\LF}[1]{{#1}^{\mathrm{LF}}}\newcommand{\Lap}[1]{{#1}^{\mathrm{L}}}\newcommand{\ds}{*\!*}\newcommand{\cond}[2]{\frac{\mathrm{D} #1}{\mathrm{D} #2}}\newcommand{\pard}[2]{\frac{\partial #1}{\partial #2}}\newcommand{\totd}[2]{\frac{\mathrm{d}#1}{\mathrm{d}#2}}\newcommand{\pardd}[3]{\frac{\partial^2 #1}{\partial #2 \partial #3}}\newcommand{\Rey}{\mbox{Re}}\newcommand{\Imag}{\mbox{Im}}\newcommand{\Fpint}{=\!\!\!\!\!\!\!\int}\newcommand{\txi}{\tilde\xi}\newcommand{\dxi}{\delta\xi}\newcommand{\tpsi}{\tilde\psi}\newcommand{\dpsi}{\delta\psi}
\makeatletter
\def\sgn{\mathop{\operator@font sgn}}

\makeatother

\title{Spontaneous autophoretic motion of isotropic particles}

%\title{Theory of autophoretic swimming for isotropic particles}
\author{S\'ebastien Michelin}
\email{sebastien.michelin@ladhyx.polytechnique.fr}
\affiliation{LadHyX -- D\'epartement de M\'ecanique, Ecole
 polytechnique, 91128 Palaiseau Cedex, France}
\author{Eric Lauga}
%\email{elauga@ucsd.edu}
\affiliation{Department of Mechanical and Aerospace Engineering, University of California San Diego, 9500 Gilman Drive, La Jolla CA 92093, USA.}
\author{Denis Bartolo}
%\email{denis.bartolo@ens-lyon.fr}
\affiliation{Laboratoire de Physique, Ecole Normale Sup\'erieure de Lyon, Universit\'e de Lyon and CNRS, 46, all\'ee d'Italie, F-69007 Lyon, France}
%\affiliation{PMMH, CNRS UMR7636, ESPCI ParisTech, Universit\'e Pierre et Marie Curie, Universit\'e Paris Diderot, 10, rue Vauquelin, 75231 Paris Cedex 05, France}

\date{\today}
\begin{abstract}

 Suspended colloidal particles interacting chemically with a solute can self-propel by autophoretic motion when they are asymmetrically patterned (Janus colloids).  Here we demonstrate theoretically  that such anisotropy is not necessary for locomotion and that the nonlinear interplay between surface osmotic flows  and solute advection can produce spontaneous, and self-sustained motion of isotropic particles. Solving the classical autophoretic framework for isotropic particles, we show that, for given material properties, there exists a critical particle size (or P\'eclet number) above which spontaneous symmetry-breaking and autophoretic motion occur.  A hierarchy of instabilities is further identified for quantized critical P\'eclet numbers.

%Suspended colloidal particles interacting chemically with a solute are able to self-propel by autophoretic motion when they are asymmetrically patterned (Janus colloids).  Here we demonstrate that the chemical anisotropy is not a necessary condition to achieve locomotion. The non linear interplay between surface osmotic flows and solute advection can produce spontaneous, and self-sustained motion of isotropic particles. We solve, for  a spherical particle, the classical nonlinear autophoretic theoretical framework at arbitrary P\'eclet number. For a given set of material parameters, there  exists a critical particle size, or equivalently a critical P\'eclet number,  above which spontaneous autophoretic motion occurs.  The  flow  induced by the particle  further displays a hierarchy of instabilities  associated with quantized critical P\'eclet numbers.   Using   numerical solutions of the full (unsteady) diffusiophoretic problem we confirm our analytical  predictions and show that, above the instability threshold,  the isotropic particles reach a steady  swimming state with broken front-back symmetry in the concentration field and the hydrodynamic signature of a ``pusher'' swimmer. This instability to propulsion could be relevant to the high-throughput production of self-propelled particles. 
\end{abstract}
%\pacs{82.70.Dd, 47.70.Fw, 47.20.-k, 87.85.gj}
%82.70.Dd Colloids
%47.70.Fw Chemically reactive flow
%47.20.-k	Flow instabilities
%87.85.gj movement and locomotion, 
\maketitle

The locomotion of microorganisms has long been used as  a motivation and a practical inspiration for the design of synthetic self-propelled particles. Typically, biological cells  generate  propulsion by deforming their slender appendages, termed flagella or cilia, in a non-time-reversible fashion~\cite{Lauga:2009ul}. However,  and perhaps not surprisingly given  the numerous microfabrication challenges,  no genuinely self-propelled micro-swimmer  has been manufactured  in the lab so far. Instead, man-made biomimetic propellers are  driven by external  torques or forces.  That actuation allows either to deform soft propellers whose deformed shape induce propulsion \cite{flexible1, flexible2,OnshunSoftMatter}, to continuously generate  propulsion in chiral shapes \cite{GhoshFischer09,NelsonCargo}, or to exploit interactions with  surfaces \cite{surface1, surface2, surface3}.

An alternative route for the production of artificial small-scale swimmers has proven to be much more successful. It consists in making miniaturized chemically powered ``engines'' with no moving parts,  typically made of reactive Janus beads or rods~\cite{Howse:2007ed,Ebbens:2010fd,Zhao:2012ch}. The reaction products  released by these chemically-asymmetric particles create  concentration gradients which induce a net phoretic fluid motion near their surface leading to locomotion.  Theoretically, the   interest in these so-called autophoretic swimmers was triggered by a  theoretical model which accounted for such a novel propulsion mechanism in a simple and generic fashion \cite{Golestanian:2005el,Golestanian:2007hu}. This model was then  further elaborated to include the nonlinear interplay between the colloid motion and the advection of the reactants~\cite{CordovaFigueroa:2008db,BRADY:2010bh,Julicher:2009hw} or a more complex kinetic route for the surface chemistry \cite{ebbens2012}, to deal with the rotational Brownian motion of the swimmers~\cite{Golestanian:2009io}, and to detail the microscopic coupling between the concentration gradients and the fluid flows at small scales~\cite{Sabass:2012ct}.  

In order to self-propel, autophoretic swimmers are chemically patterned, and it is  the asymmetries in the chemical reactions on their surfaces which are responsible for locomotion in the first place. This requirement would make it thus difficult to achieve high-throughput production.  An ingenious solution to such an engineering issue was recently offered with the production of isotropic  self-propelled Marangoni droplets~\cite{Thutupalli:2011bv}.  In a mechanism  akin to the one responsible for the spontaneous motion of reactive droplets surfing on fluid interfaces~\cite{Sumino:2005ck,Toyota:2009cg},  a net flow is induced around the  droplets by interfacial stresses arising from the self-generated gradients of reactive surfactants~\cite{yoshinaga2012,Schmid:2000js,Thutupalli:2011bv}. Unlike autophoretic Janus particles which swim as a result of  built-in design asymmetries, the concentration of surfactant molecules at the surface of  reactive droplets is spontaneously broken, leading to propulsion.

In this Letter, we demonstrate that asymmetry in the shape or the chemistry of the  colloidal particle, assumed in most existing studies on autophoretic swimmers \cite{Golestanian:2005el,Golestanian:2007hu,CordovaFigueroa:2008db,BRADY:2010bh,ebbens2012}, is not necessary for locomotion. We show that  isotropic spherical particles can   achieve self-propulsion through a spontaneous symmetry-breaking mechanism, a route to locomotion akin to that of a larger variety of  biological systems \cite{vandergucht2009}.
We solve analytically the classical nonlinear autophoretic theoretical framework at arbitrary P\'eclet number. 
We show that, for a given set of material properties, there  exists a critical particle size above which the non-swimming purely diffusive state is unstable and spontaneous autophoretic motion occurs.  In addition,   the  flow  induced by the reactive particles displays a hierarchy of instabilities associated with quantized critical P\'eclet numbers.  Using   numerical solutions of the full unsteady diffusiophoretic problem we confirm our analytical  predictions and show that, above the instability threshold, isolated isotropic particles reach, after a transient,  a  steady  swimming state with broken front-back symmetry in the concentration field and the hydrodynamic signature of a ``pusher'' swimmer.

We consider a spherical particle of radius $a$ located in a Newtonian fluid of shear viscosity $\eta$. We focus on the limit of steady Stokes flow so that the inertia of both the fluid and the particle is negligible. The sphere is chemically active and  either  emits  or captures  a single type of solute particles with an isotropic surface emission rate denoted $\cal A$~\cite{Golestanian:2005el, Golestanian:2007hu}. Within the standard minimal framework, we ignore in what follows the  specifics of the chemical reaction  occurring at its surface, and all  other reactants and reaction products, which are assumed to interact only weakly with the particle. The solute interacts with the spherical particle through a short-range  potential \change{ and we focus on the classical limit $\lambda\ll a$, with $\lambda$ the characteristic potential range}~\cite{anderson1989}. %The competition between the  interaction potential and  molecular diffusion sets the thickness of the so-called diffuse layer, $\lambda$. 
%As in classical approaches, we consider the  \change{limit $\lambda\ll a$, where $\lambda$ is the characteristic potential range}~\cite{anderson1989}. % thin-diffuse layer approximation, $\lambda\ll a$\change{, with $\lambda$ the characteristic interaction range}~\cite{anderson1989}. 

Tangential gradients in solute concentration induce  a net slip (phoretic) velocity  outside the diffuse layer,  together with a tangential (Marangoni) stress discontinuity at the particle surface~\cite{anderson1989,Julicher:2009hw}. For a solid particle,  Marangoni stresses vanish, and the phoretic slip velocity is set by the local chemical-potential gradient parallel to the surface~\cite{Prieve:1987cu}. For small variations of the concentration, $c({\bf r},t)$, its azimuthal component is given by\begin{equation}
u_\theta(r=a)=\frac{\mathcal{M}}{a}\pard{c}{\theta}\cdot
\label{utheta}
\end{equation}
The mobility, $\mathcal{M}\sim {\pm k_BT\lambda^2}/{\eta}$,   
can either be positive or negative depending on the form of the solute-surface interactions ($k_B$ is the Boltzmann constant, and $T$ the temperature) \cite{anderson1989}. 
Assuming that the  solute has a molecular diffusivity ${\cal D}$, 
the typical autophoretic velocity  which  sets the swimming speed of a Janus colloid is  ${\cal V}=|{\cal AM}|/{\cal D}$ \cite{Golestanian:2007hu}. Non-dimensionalizing   velocities,  lengths, and concentrations in the transport equations by ${\cal V}$,   $a$, and $a {|\mathcal A|} /{\cal D}$  respectively,   the  coupled fluid flow- solute advection-diffusion  problem  takes the  form
\begin{align}
\nabla^2\ub&=\nabla p,\qquad \nabla\cdot \ub=0\label{eq:phor1},\\
|\Pe|&\left(\pard{c}{t}+\ub\cdot\nabla c\right)=\nabla^2 c, \label{eq:phor1-2}
\end{align}
where we have defined the (signed)  P\'eclet number as 
$\Pe= {\mathcal{A}\mathcal{M}a}/{{\cal D}^2}$. 
We henceforth  consider axisymmetric solutions only, so that in spherical polar coordinates the solute concentration is written as $c(r,\mu,t)$ with $\mu\equiv\cos\theta$, and only rectilinear motion along the axis of symmetry $\eb_z$ is considered.  In dimensionless units, the surface activity and the particle mobility become unitary numbers,  $A={\cal A}/{|\cal A|}=\pm 1$, $M={\cal  M}/{|\cal M|}=\pm1$, and the boundary conditions  \change{in the far-field} and on the particle surface become 
\begin{align}
\label{swim} \ub(r\rightarrow\infty)&=-U\eb_z,\quad c(r\rightarrow\infty)=c_\infty,\\
\pard{c}{r}(r=1)&=- A,\quad u_r(r=1)=0,\\
 u_\theta(r=1)&=- { M}\sqrt{1-\mu^2}\pard{c}{\mu}(r=1).\label{eq:phor2}
\end{align}
In Eq.~\eqref{swim}, $U(t)$ is the dimensionless swimming velocity,  obtained  from the  force-free condition. Using  the reciprocal  theorem,  $U(t)$ is found to be the surface average of the slip velocity~\cite{Stone:1996vd, Golestanian:2007hu} and, using Eq.~\eqref{utheta},  is written  in terms of the first moment of $c(1,\mu,t)$  as
\begin{equation}
 U(t)=- {M}\int_{-1}^1\mu \,c(1,\mu,t)\,\dd\mu.
 \label{particlevelocity}
\end{equation}
The dimensionless problem is then fully characterized by the signs of both $A$ and $M$,  the   far-field concentration, $c_\infty$, and the value of $|\Pe|$.

In the  case of uniform surface activity (constant value of $ A$), a trivial solution exists at all  $\Pe$ numbers, namely the solute concentration is isotropic, $\bar{c}=A/r+c_\infty$, leading to no net flow ($\bar\ub=0$) and zero swimming velocity ($\bar{U}=0$).  However, both the solute transport equation, Eq.~\eqref{eq:phor1-2},  and the boundary conditions  on the particle, Eq.~\eqref{eq:phor2}, couple  the swimming problem with the solute dynamics.  A small fluctuation of the particle velocity would result in a polar perturbation of the flow field.  Due to the nonlinear convective coupling, ${\bf u}\cdot\nabla c$, this velocity fluctuation would lead to  a polarization of the concentration field  around the finite-size particle.  In turn, the first moment of the surface concentration, $c(1,\mu,t)$, would become finite and, depending on its sign in relation to the initial perturbation, increase or decrease  the particle velocity through Eq.~\eqref{particlevelocity}. If the velocity  decreased, the initial perturbation would be stabilized, and no net motion could occur as a result of an infinitesimal fluctuation. However, if the velocity  increased, the broken  symmetry in the solute concentration would be amplified, and spontaneous motion would occur. 

To quantify  the  conditions for  spontaneous motion, we analytically investigate the stability of the isotropic state. Defining $c=\bar{c}+c'$, $\ub=\ub'$, $U=U'$,  and subsequently dropping the primes to denote perturbations, the Stokes flow problem around the sphere can be solved analytically using the \change{so-called squirming modes decomposition \cite{blake1971,michelin2011}. The streamfunction $\psi$ and solute concentration $c$ are decomposed azimuthally onto orthogonal modes
\begin{equation}
\psi(r,\mu,t)=\sum_{n=1}^\infty\frac{2n+1}{n(n+1)}\alpha_n(t)\psi_n(r)(1-\mu^2)L_n'(\mu),\quad c(r,\mu,t)=\sum_{n=0}^\infty c_n(r,t)L_n(\mu),
\end{equation}
with $\psi_1(r)=(1-r^3)/3r$, $\psi_n(r)=(r^{-n}-r^{-n+2})/2$ for $n\geq 2$, and $L_n(\mu)$ the $n$-th Legendre polynomial. The squirming mode intensities, $\alpha_n(t)$, are obtained directly from the slip velocity on the particle surface~\cite{michelin2011}, as
\begin{equation}
\alpha_n(t)=\frac{1}{2}\int_{-1}^1\sqrt{1-\mu^2}L_n'(\mu)u_\theta(r=1,\mu,t)\dd\mu.
\end{equation}
The first mode is the only one contributing to the swimming velocity ($\alpha_1(t)=U(t)$), and is therefore termed the swimming mode. The second mode corresponds to the flow created by a stresslet (i.e. a force dipole)~\cite{Batchelor_Stress70}. Higher order modes correspond to higher order singularities decaying faster  in the far field.

Because the trivial solution is isotropic, we can project both the advection-diffusion equation, Eq.~\eqref{eq:phor1-2}, and the  definition of the slip velocity, Eq.~\eqref{eq:phor2}, along these modes, and obtain a set of independent problems for $\{\alpha_n(t), c_n(r,t)\}$. For $n=1$,} this leads to the following inhomogeneous eigenvalue problem for the first moment of the concentration profile, $ c_1(r,t)$: 
\begin{align}
|\Pe|\pard{c_1}{t}-&\frac{1}{r^2}\left[\pard{}{r}\left(r^2\pard{c_1}{r}\right)-2c_1\right]=\frac{U\Pe}{\change{M} r^2}\left(\frac{1}{r^3}-1\right),\label{eq:perturb1}\\
U(t)=-\frac{2}{3}&{ M}c_1(1,t),\quad c'_1(1,t)=0,\quad c_1(r\rightarrow\infty,t)\sim 0.\label{eq:perturb2}
\end{align}
Note that the first moment, $c_1$, evaluated at $r=1$ is proportional to the instantaneous  swimming speed of the particle, (Eq.~\ref{eq:perturb2}). Looking for eigenmodes of the form $c_1(r,t)=\ee^{\sigma t} c(r)$, it can be shown that all eigenvalues $\sigma$ of Eqs.~\eqref{eq:perturb1}--\eqref{eq:perturb2} are real, and that any $\sigma<0$ is a solution. We focus here exclusively on potential unstable modes ($\sigma>0$) and define $\beta\equiv\sqrt{\sigma |\Pe|}>0$. Introducing the rescaled radial variable $x\equiv\beta r$, the  function $C(x)\equiv c(\beta x)$ satisfies 
\begin{equation}
\totd{}{x}\left(x^2\totd{C}{x}\right)-\left(2+x^2\right) C=\frac{2\Pe}{3}\left(\frac{\beta^3}{x^3}-1\right).\label{above}
\end{equation}
The general solution of Eq.~\eqref{above} satisfying the  far-field condition $C(x\rightarrow\infty)=0$ is 
\begin{align}
C(x)=&\frac{2\Pe}{3}\left\{\frac{1}{x^2}+\beta^3\left[\frac{A(x)}{8x^2}+\frac{B(x)}{8x}+\frac{1}{4x^3}\right]\right\}+\frac{b\ee^{-x}(1+x)}{x^2}
\end{align}
where $b$ is an integration constant to be determined, and  $A(x)=\sinh(x)\textrm{Chi}(x)-\cosh(x)\textrm{Shi}(x)$, and 
$B(x)=\sinh(x)\textrm{Shi}(x)-\cosh(x)\textrm{Chi}(x)$, with $\textrm{Chi}(x)$ and $\textrm{Shi}(x)$ being the hyperbolic cosine and sine integral functions respectively~\cite{abramowitz1964}. Applying  the two boundary conditions on the sphere, $C'(\beta)=0$, and $C(\beta)=1$, and using the definitions of $A(x)$ and $B(x)$ yields a final implicit expression for the growth rate, $\sigma=\beta^2/ |\Pe| $, as a function of the signed P\'eclet number
\begin{equation}
\Pe=\frac{12\beta^2+24\beta+24}{\displaystyle\beta^4 \int_\beta^\infty\frac{\ee^{\beta-t}}{t}\dd t+6+\beta^2-\beta^3-2\beta}\cdot
\label{final}
\end{equation}
%%%
\begin{figure}
\begin{center}
\includegraphics[width=0.58\columnwidth]{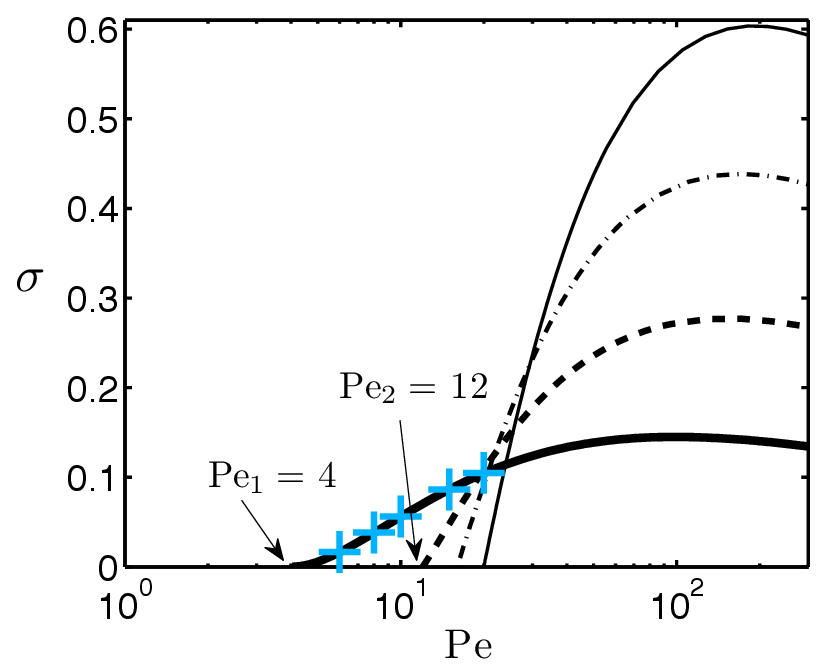}
\caption{(Color online) Growth rate of the unstable swimming  mode (theoretical prediction, solid line) as a function of the P\'eclet number, $\Pe$, for emitting particles  and positive mobility ($A= M =1$). Spontaneous symmetry-breaking of the concentration field  and swimming occur at  $\Pe = 4$. The crosses represent the  growth rate of the swimming mode $n=1$ as obtained from numerical simulations of the full unsteady  problem.  The growth rate of the unstable modes of azimuthal order $n=2,$ 3 and 4 are also shown (theoretical prediction, dashed, dash-dotted and thin-solid, respectively). Note that the hierarchy in the growth rates is reversed at high Pe.}\label{fig:growthrate}
\end{center}
\end{figure}

For positive values of $\beta$, the right-hand side of Eq.~\eqref{final} is strictly greater than $4$.  For any value of $\Pe$ above this critical value, a single positive value of $\beta$ exists such that  Eq.~\eqref{final} is satisfied.  Consequently, the fluctuations of the first moment, $c_1$, and the particle swimming speed, $U$, are exponentially amplified if the  condition $\Pe\geq\Pe_{1}=4$ is satisfied. Given that $\Pe$ is a signed quantity, the instability condition requires that $ MA = 1$. Particles with positive (resp.~negative) mobility $ \cal M$ are unstable only if they  have a positive (resp.~negative) flux $  \cal  A$ corresponding to the case of emitting (resp.~absorbing) the solute  on the particle surface.
 The growth rate of the unstable swimming mode is shown as a function of $\Pe$ in Fig.~\ref{fig:growthrate} (solid line). 
 \change{The existence of a critical Peclet number implies that there exists a critical particle radius above which an isotropic reactive particle  would undergo spontaneous motion. Recently, 
a study of light-activated colloids reported the spontaneous surfing motion of reactive isotropic colloids lying on a solid surface \cite{palacci2013}, a system which might be  a potential candidate to test, at least qualitatively, our predictions.}
 
 %From a practical point,  the existence of a critical Peclet number implies that   there exists a critical particle radius,  $a_{1}=4{\cal D}/{\cal V}$, above which an isotropic reactive particle at rest  would undergo spontaneous motion. For instance, considering the prototypal example of a Pt-coated bead in a $\rm H_2O/H_2O_2$ solution, we have $ {\cal V}\approx 20\,\mu\rm m\,s^{-1}$ \cite{Golestanian:2007hu,Howse:2007ed,Palacci:2010hk} and  taking  ${\cal D}\approx10^{-9}\,{\rm m^2/s}$~\cite{Kern:1954ux} we obtain a critical radius   $a_{1}\approx 100\,\mu\rm m$, which is  experimentally accessible.  This critical particle radius could   be further  reduced either by increasing the fluid viscosity, or by using  reactants molecules with a larger molecular weight~\cite{Ebbens:2010fd}. 

In order to further investigate the possibility for long-time self-propulsion, we need to go beyond the above linear stability analysis and check whether the  asymptotic state of the concentration field is  compatible with  swimming.  In order to do so, numerical simulations of the full unsteady diffusiophoretic problem, Eqs.~\eqref{eq:phor1}--\eqref{eq:phor2}, are performed: the Stokes flow problem is solved explicitly using the squirming mode decomposition, and the advection-diffusion problem is marched in time using a semi-explicit scheme, finite differences in the radial directions, and the Legendre spectral decomposition for the azimuthal dependence \cite{michelin2012a}. A small velocity perturbation is imposed on the spherical particle initially at rest. Regardless of the amplitude of the initial perturbation, the system is stable and returns to its initial rest state after perturbation for $\Pe<4$. In contrast, when $\Pe > 4$, the swimming velocity of the particle grows exponentially, and the growth rates obtained numerically are   in quantitative agreement  with our analytic predictions, see Fig.~\ref{fig:growthrate}.

\begin{figure}
\begin{center}
\includegraphics[width=.68\columnwidth]{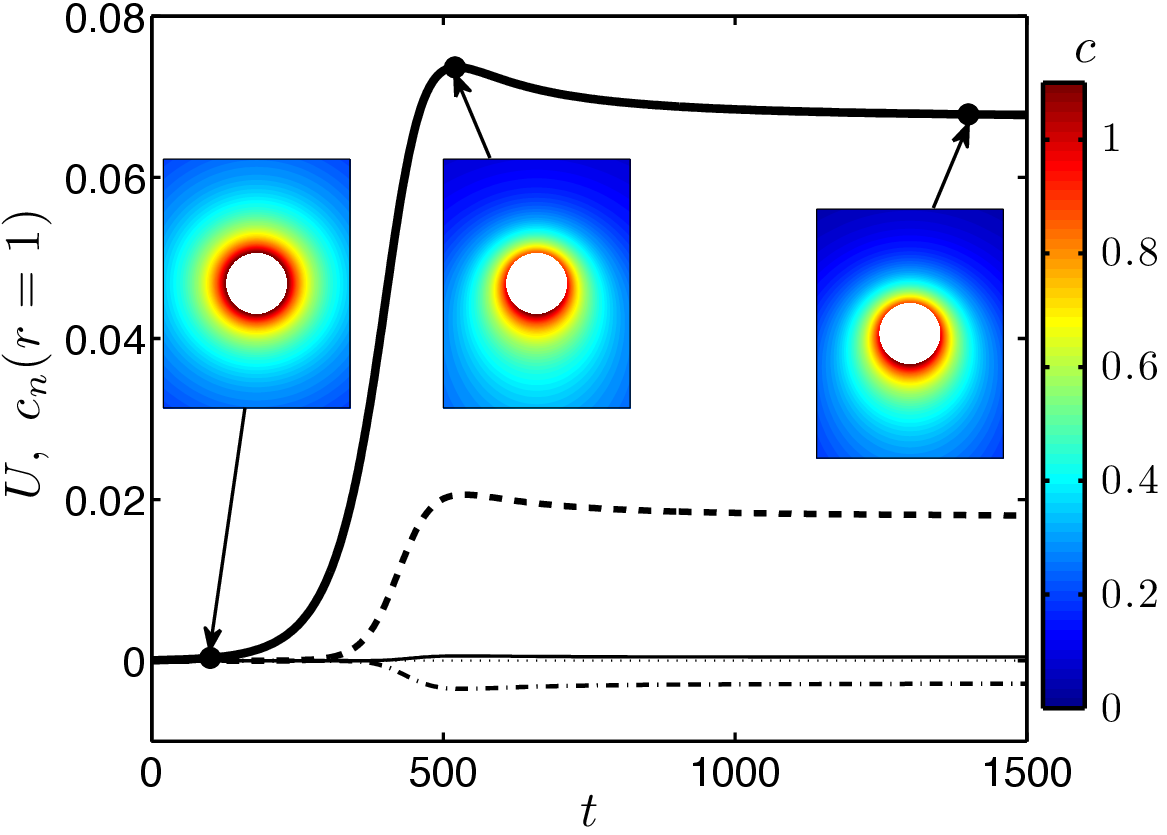}
\caption{(Color online) Time-evolution of the instantaneous swimming velocity, $U(t)$ (thick solid line), and of three  higher squirming modes, $c_n(r=1,t)$:  mode $n=2$, which  corresponds to the magnitude of the  stresslet  (dashed line), $n=3$ (dash-dotted line) and $n=4$ (thin solid line). Results are displayed for $\Pe=6$ and ${A}{M}=1$. The solute concentration is shown at three different times revealing the establishment of a front-back asymmetry for an upward swimming motion (see Supplementary video)}\label{fig:movie}
\end{center}
\end{figure}

\begin{figure}
\begin{center}
\includegraphics[width=.7\columnwidth]{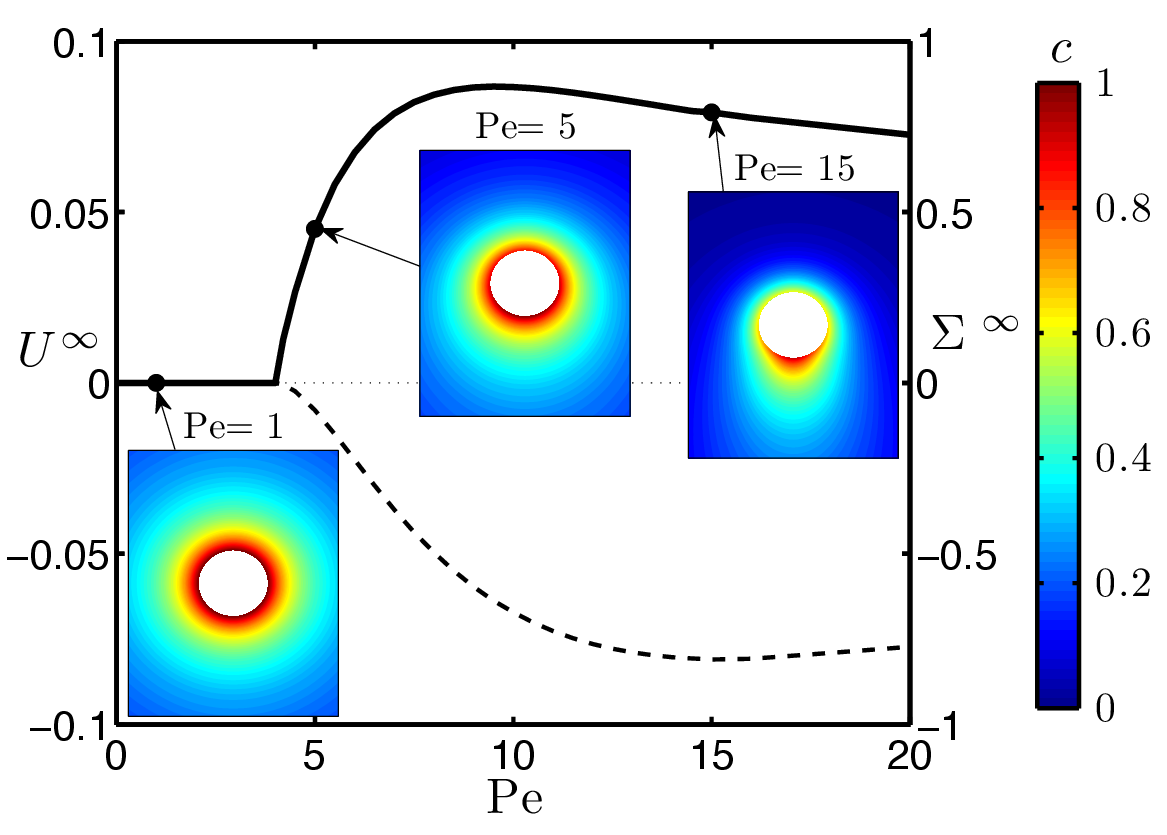}
\caption{(Color online) Long-time spontaneous swimming velocity, $U^\infty$ (solid line), and magnitude of the long-time induced stresslet,   $\Sigma^\infty\equiv-4\pi M c_2(r=1,t=\infty)$ (dashed line), as a function of the P\'eclet number (with ${A}{M}=1$). The steady state solute distribution around the particle is shown for $\Pe=1$, $\Pe=5$, and $\Pe=15$ for an upward swimming motion. The swimmer is always a ``pusher'' in the far field.}\label{fig:swim}
\end{center}
\end{figure}

Turning to the long-time behavior, our computations confirm that the  asymptotic state of the concentration field is compatible with locomotion. This is illustrated  in Fig.~\ref{fig:movie} for $\Pe=6$, where we plot the time-evolution of the instantaneous particle swimming speed (thick solid line).  \change{Three typical snapshots in Fig.~\ref{fig:movie} (see also the associated video)} illustrate the corresponding evolution of the solute concentration field. A steady state is reached and the swimming speed plateaus to a finite  value in the long-time limit (\change{note the surprising existence of a local maximum velocity  during the transient dynamics}).  Repeating the simulations for a  range of P\'eclet numbers, we plot in   Fig.~\ref{fig:swim} the nonlinear variation of the long-time swimming velocity, $U^\infty$,  as a function of $\Pe$. Figure~\ref{fig:swim}  demonstrates the supercritical nature of the autophoretic instability: for $\Pe>4$, any infinitesimal perturbation of the isotropic state will lead to spontaneous self-propulsion. Note the non-monotonic variation with an optimal value of $\Pe\approx 9$ leading to the highest asymptotic swimming speed.

% For a given set of material parameters, and unlike autophoretic Janus particles~\cite{Golestanian:2007hu}, the swimming speed here explicitly  depends on the particle radius $a\propto \Pe$. Physically, this difference arises because  the concentration gradient around an isotropic particle is not merely set by the particle size but instead by a dynamical scale.

So far we focused exclusively on \change{unstable swimming modes. In order to address the collective dynamics of such particles, it is necessary to consider higher order squirming modes}. %one would need to gain insight on the associated flow field away from the particle. The higher-order squirming modes are defined as $c_n(r,t)=({n+1}/{2})\int_{-1}^1c(r,\mu,t)L_n(\mu)\dd\mu$, where $L_n(\mu)$ is the Legendre polynomial of order $n$. 
%As discussed above, $c_1$ is the (only) swimming mode. The flow field associated with the mode $c_2$ is the one with the slowest spatial decay, that of a stresslet (i.e. of a force dipole)~\cite{Batchelor_Stress70}. Higher  values of $n$ correspond to higher-order flow singularities in the far field. 
Repeating, for $ n \geq 2$,  the theoretical  approach presented above for $n=1$, we  can obtain  the unstable growth rates of each mode as a function of $\Pe$. The results reveal  a hierarchy of supercritical instabilities corresponding to an infinite set of quantized critical P\'eclet numbers,  $\Pe_{n}=4(n+1)$, for mode $n$. This hierarchy of instabilities is illustrated in Fig.~\ref{fig:growthrate} where we plot the  dependence of the  growth rates for modes 2, 3, and 4 on the P\'eclet number. Notably, the stresslet mode becomes unstable at $\Pe_2=12$.

Beyond linear analysis, the saturation of the swimming velocity results from the nonlinear evolution of  concentration fluctuations having a $n=1$ symmetry into higher-order  $n$-modes, as shown in  Fig.~\ref{fig:movie}. Although the P\'eclet number is below the critical value for all modes but the first one to be unstable ($\Pe=6$ in Fig.~\ref{fig:movie}), the nonlinear dynamics leads non-zero  values for the other modes (modes 2, 3, and 4 are shown in Fig.~\ref{fig:movie}). In particular, the long-time value of the induced stresslet, $\Sigma^\infty$, is shown in Fig.~\ref{fig:swim}. Independently of the signs of both  $\cal A$ and $\cal M$, the unstable particle  induces in the far field a flow with the symmetry of a ``pusher" swimmer, similarly to flagellated bacteria \cite{Lauga:2009ul}.

Similarly to Marangoni flows which have been observed for over 100 years to trigger the self-propulsion of camphor boats floating on water, we demonstrated in this letter that  self-phoretic flows past isotropic particles show an instability to a spontaneous swimming state. The phenomenon discovered here could be exploited to readily design  self-propelled ``submarines" out of isotropic colloidal particles.  Although our description of the surface chemistry is particularly simple  (a fixed-rate absorption or release of solute), the instability and self-propulsion mechanism  remain in fact valid for a much broader class of surface  chemistry (see supplementary material\footnote{See supplementary material at AIP TO ADD URL that demonstrates that the spontaneous motion is a generic mechanism which applies to more realistic reaction kinetics.}). These results are expected to be applicable generically to particles of different shapes or interacting with many chemical species, and suggest a simple experimental model system to carry out physical studies of  active systems. 
 
We acknowledge valuable discussions with Olivier Dauchot and John Brady. This work was supported in part by the NSF through grant number CBET-0746285 (EL).

%\bibliography{refs_phoretic}
%\bibliographystyle{apsrev-title}
%\end{document}

\end{document}